\documentclass{article}

\usepackage{arxiv}

\usepackage[utf8]{inputenc} 
\usepackage[T1]{fontenc}    
\usepackage{hyperref}       
\usepackage{url}            
\usepackage{booktabs}       
\usepackage{amsfonts}       
\usepackage{nicefrac}       
\usepackage{microtype}      
\usepackage{lipsum}
\usepackage{fancyhdr}       
\usepackage{graphicx}       
\graphicspath{{media/}}     
\usepackage{cite}
\usepackage{amsmath,amssymb,amsfonts}
\usepackage{algorithmic}
\usepackage{graphicx}
\usepackage{algorithm,algorithmic}
\usepackage{hyperref}
\hypersetup{hidelinks=true}
\usepackage{textcomp}
\usepackage{algorithm}
\usepackage{mathtools}
\usepackage{multirow}
\usepackage{longtable, booktabs}

\title{A Physics-Informed Low-Shot Learning For sEMG-Based Estimation of Muscle Force and Joint Kinematics}

\author{ \href{https://orcid.org/0000-0001-8424-6996}{\includegraphics[scale=0.06]{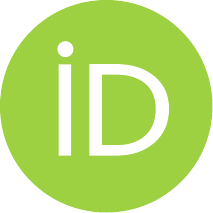}\hspace{1mm}Yue~Shi}\\
	School of Electronic and Electrical Engineering, \\
	The University of Leeds, \\
	Leeds, UK \\
	\texttt{y.shi1@leeds.ac.uk} \\
	\And
	\href{https://orcid.org/0000-0000-0000-0000}{\includegraphics[scale=0.06]{orcid.pdf}\hspace{1mm}Shuhao~Ma} \\
	School of Electronic and Electrical Engineering, \\
	The University of Leeds, \\
	Leeds, UK \\
	\texttt{elsma@leeds.ac.uk} 
    \And
	\href{https://orcid.org/0000-0000-0000-0000}{\includegraphics[scale=0.06]{orcid.pdf}\hspace{1mm}Yihui~Zhao} \\
	Bristol robotic lab, \\
    University of Bristol, \\
    Bristol, UK\\
    \texttt{yihui.zhao@bristol.ac.uk} 
    \And
	\href{https://orcid.org/0000-0000-0000-0000}{\includegraphics[scale=0.06]{orcid.pdf}\hspace{1mm}Zhiqiang~Zhang} \\
	School of Electronic and Electrical Engineering, \\
	The University of Leeds, \\
	Leeds, UK \\
	\texttt{z.zhang3@leeds.ac.uk} 
    \And
	\href{https://orcid.org/0000-0000-0000-0000}{\includegraphics[scale=0.06]{orcid.pdf}\hspace{1mm}Zhiqiang~Zhang} \\
	School of Electronic and Electrical Engineering, \\
	The University of Leeds, \\
	Leeds, UK \\
}

\renewcommand{\shorttitle}{\textit{arXiv} Template}

\hypersetup{
pdftitle={A template for the arxiv style},
pdfsubject={q-bio.NC, q-bio.QM},
pdfauthor={Yue~Shi},
pdfkeywords={muscle force and joint kinematics, surface Electromyographic, low-shot learning, generative adversarial network, physics-informed optimization, mode collapse},
}

\begin{document}
\maketitle

\begin{abstract}
Muscle force and joint kinematics estimation from surface electromyography (sEMG) are essential for real-time biomechanical analysis of the dynamic interplay among neural muscle stimulation, muscle dynamics, and kinetics. Recent advances in deep neural networks (DNNs) have shown the potential to improve biomechanical analysis in a fully automated and reproducible manner. However, the small sample nature and physical interpretability of biomechanical analysis limit the applications of DNNs. 
This paper presents a novel physics-informed low-shot learning method for sEMG-based estimation of muscle force and joint kinematics. This method seamlessly integrates Lagrange's equation of motion and inverse dynamic muscle model into the generative adversarial network (GAN) framework for structured feature decoding and extrapolated estimation from the small sample data. Specifically, Lagrange's equation of motion is introduced into the generative model to restrain the structured decoding of the high-level features following the laws of physics. And a physics-informed policy gradient is designed to improve the adversarial learning efficiency by rewarding the consistent physical representation of the extrapolated estimations and the physical references. 
Experimental validations are conducted on two scenarios (i.e. the walking trials and wrist motion trials). Results indicate that the estimations of the muscle forces and joint kinematics are unbiased compared to the physics-based inverse dynamics, which outperforms the selected benchmark methods, including physics-informed convolution neural network (PI-CNN), vallina generative adversarial network (GAN), and multi-layer extreme learning machine (ML-ELM). 
\end{abstract}


\section{Introduction}
\label{sec:introduction}
Human movements involve complex interactions within the neuromuscular system. The surface electromyography (sEMG)-driven estimation of muscle force and joint kinematics dynamics provides detailed biomechanical analysis to understand the neuromuscular system \cite{falisse2019rapid, modenese2020automated}, which benefits various applications, such as sports rehabilitation treatments \cite{smith2021review, kotsifaki2022single},  \cite{arones2020musculoskeletal}, and optimizing robotic design for individuals with impairments \cite{grabke2019lower}. Although physics-based models explicitly explain and map sEMG signals to joint kinematics, the high cost of their static optimization has always limited the practical applications of these models \cite{zhang2022physics, zhao2022computational}. \par

Recently, deep neural networks (DNNs) provide an alternative solution to map the sEMG signals to the joint kinetics and kinematics \cite{dorschky2020cnn, johnson2018predicting}. In this kind of model, the multi-layer convolution architecture has been explored to establish relationships between movement variables and neuromuscular status \cite{chaudhary2022single, zhang2022boosting}. For example, Nasr et al \cite{nasr2021musclenet} mapped the sEMG signals to the regression of joint angle, joint velocity, joint acceleration, joint torque, and activation torque, illustrating that the multi-layer convolution operators are capable of extracting underlying motor control information. Zhang et al \cite{zhang2022boosting} developed an active deep convolutional neural network to enhance the dynamic tracking capability of the musculoskeletal model on unseen data.  \par

Despite the advantages, traditional DNNs are data-hungry and their performance is highly dependent on the quantity and quality of data \cite{solares2020deep}. Meanwhile, biomechanics analysis is typically a physics-based extrapolation process with small sample nature \cite{shadlen1996computational, holder2020systematic}. Therefore, it is a challenge to train DNNs with small sample data so that the DNNs perform consistently with the physics-based model. To fill this research gap, the low-shot learning (LSL) technique has attracted many researchers' attention \cite{hu2022can, tam2022siamese, rahimian2021trustworthy}. For example, Rahimian \textit{et al} \cite{rahimian2021fs} introduced a Few-Shot Learning Hand Gesture Recognition (FS-HGR) model to enhance the generalization capability of DNNs from a limited number of instances. Lehmler \textit{et al} \cite{lehmler2022deep} explored a low-shot learning methodology that adjusts DNNs to new users with only a small size of training data. \par

In addition, the generative adversarial network (GAN) framework has shown great potential in handling physical extrapolating and predictive problems \cite{goodfellow2014distinguishability, goodfellow2020generative, chaudhary2022single}. The GAN-based model is capable of discovering the structured patterns of the references and extrapolating the underlying data distribution characteristics during the adversarial learning process \cite{shi2022latent}. For example, Chen \textit{et al} \cite{chen2022deep} tested and evaluated the performance of the deep convolutional generative adversarial network (DCGAN) on sEMG-based data enhancement, and their results indicated that the extrapolated data is able to augment the diversity of the original data. Fahimi \textit{et al} \cite{fahimi2020generative} proposed a generative adversarial learning framework for generating artificial electroencephalogram (EEG) data to extrapolate the brain-computer interface, and their findings suggest that generated EEG augmentation can significantly improve brain-computer interface performance.

In this study, we propose a physics-informed low-shot learning method for muscle force and joint kinematics estimation from multi-channel sEMG signals. This method seamlessly integrates physics knowledge with the GAN framework for structured feature decoding and extrapolated estimation from the small sample data. Specifically, Lagrange's equation of motion is introduced into the generative model to restrain the structured decoding of the high-level features following the laws of physics. And a physics-informed policy gradient is designed to improve the adversarial learning efficiency by rewarding the consistent physical representation of the extrapolated estimations and the physical references. Results show the muscle forces and joint
kinematics estimated from the proposed method are unbiased compared to the physics-based inverse dynamics.

     
    

The remainder of this paper is organized as follows: Section \ref{meth} detailed describes the algorithm of the proposed physics-informed policy gradient for reinforcement generative adversarial learning, including the mathematics framework of the algorithm and network architectures. Section \ref{Mat_and_ex} presents the material and experimental methods. Section \ref{result} discusses the experimental results and model evaluations. and Section \ref{conclusion} presents the conclusions.

\section{Physics-informed low-shot learning method}
\label{meth}

\begin{figure}[]   
    \centering  
    \includegraphics[width=3.6 in]{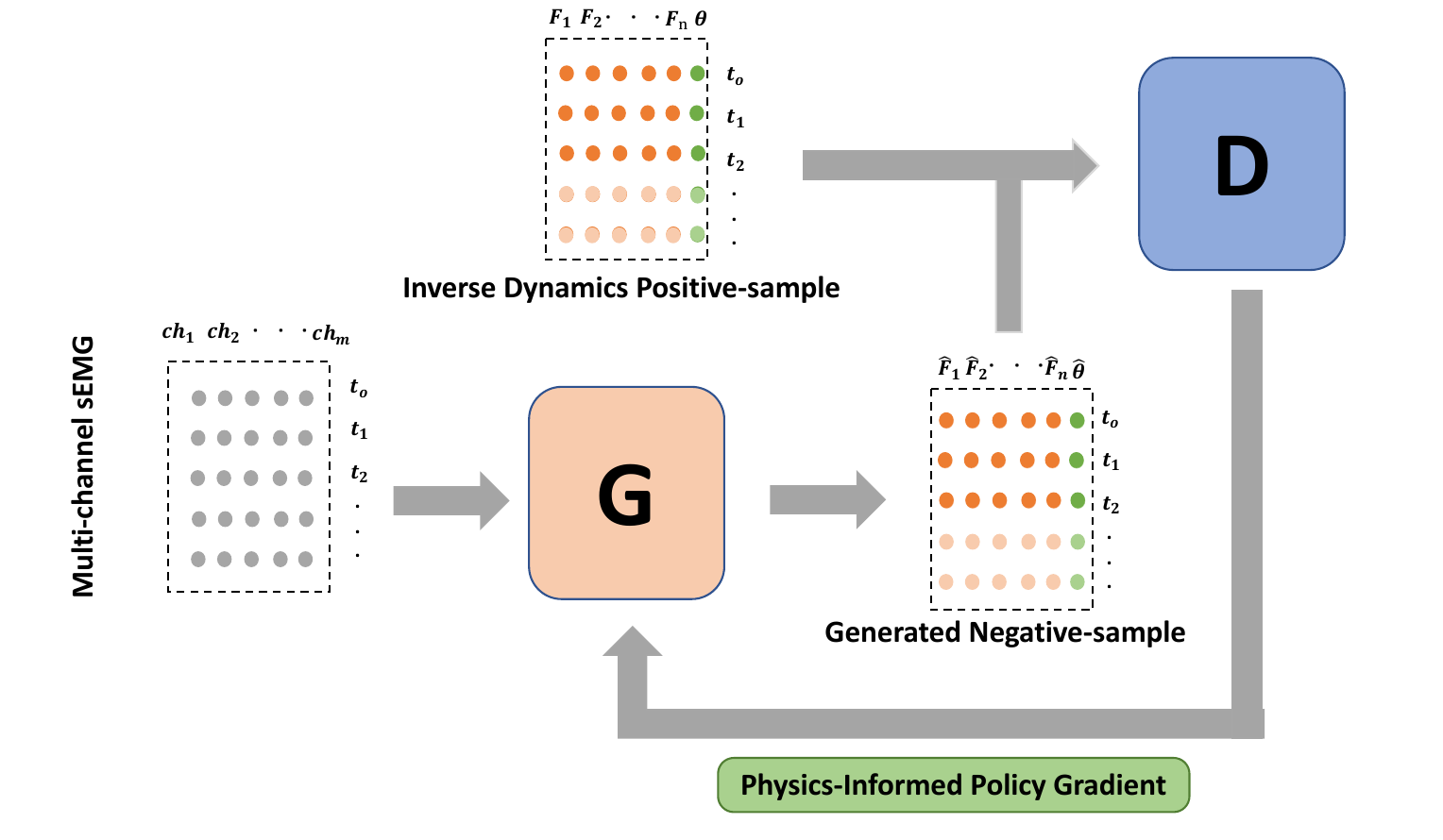}  
    \caption{The main architecture of the proposed physics-informed low-shot generative adversarial learning for muscle force and  joint kinematics prediction from multi-channel sEMG time-series} 
    \label{fig:1}  
\end{figure}

The continuous estimation of muscle forces ($F$) and joint kinematics($\theta$) from multi-channel sEMG can be denoted as the time-series generation problem. Thus, given a real multi-channel sEMG time series, we train a $\sigma$ parameterized generative network $G_{\sigma}$ to estimate the muscle force ($\hat{F}$) and joint kinematics ($\hat{\theta}$). In this section, we propose a GAN framework, as shown in Fig.\ref{fig:1}, to train the $G_{\sigma}$ on the small sample data.
Specifically, we denote the $\hat{F}$ and $\hat{\theta}$ estimated by $G_{\sigma}$ as the negative samples (see details in Section \ref{meth:4}), the ground truth ($\theta$) and the inverse dynamics-based ($F$) \cite{zhao2022musculoskeletal} as positive samples (i.e. references). The $\phi$-parameterized discriminative model $D_{\phi}$ is introduced to distinguish the positive samples and negative samples (see details in Section \ref{meth:5}). During adversarial learning, the task of $D_{\phi}$ is to determine if an input sample is positive or negative, and the task of $G_{\sigma}$ is to generate the unbiased negative samples to fool the discriminator $D_{\phi}$. The model optimization process is driven by the newly proposed physics-informed policy gradient (see details in Section \ref{meth:2}) which rewards the homogeneity of physics representation and structural characteristics between the positive and negative samples.\par

\subsection{GAN optimization via physics-informed policy gradient}
\label{meth:2}

The physics-informed policy gradient method, inspired by reinforcement learning \cite{yu2017seqgan}, aims to optimize the learning process of the GAN-based model yielding physical extrapolations from the small sample data (i.e. low-shot learning). Mathematically, the physics-informed policy gradient method maximizes its expected reward $J(\sigma)$ based on the physics law and structured characteristics from the small sample data. The $J(\sigma)$ consists of two parts, the structural reward $R_{G_{\sigma}}$ and physics representation action $Q_{D(\phi)}^{G(\sigma)}$. The $J(\sigma)$ is defined as follows.

\begin{equation}
\begin{aligned}
    J(\sigma) &= \mathbb{E}[R_{G_{\sigma}}(G_{\sigma}(sEMG_{0:T}))]\\
    & \cdot Q_{D{\phi}}^{G{\sigma}}((G_{\sigma}(sEMG_{0:T}), [F,\theta]_{0:T}) \\
    &= \mathbb{E}[R_{G_{\sigma}} ([\hat{F}, \hat{\theta}]_{0:T})] \\
    & \cdot Q_{D{\phi}}^{G{\sigma}}([\hat{F}, \hat{\theta}]_{0:T}, [F, \theta]_{0:T})
\end{aligned} 
\label{eq:1_0}
\end{equation}

\noindent where $sEMG_{0:T}$ is the input multi-channel sEMG time series for $T$ time steps. The $J(\sigma)$ is beginning with the expected reward from a predetermined state from the positive samples. And then, the $R_{G_{\sigma}}$ and $Q_{D(\phi)}^{G(\sigma)}$ will jointly optimize the generative network $G_{\sigma}$ to generate the unbiased $([\hat{F}, \hat{\theta}]_{0:T})$ following the physics laws.\par

Specifically, the structural reward $R_{G_{\sigma}}$ is computed by the $G_{\sigma}$ and defined as follows. 

\begin{equation}
\begin{aligned}
    R_G(([\hat{F}, \hat{\theta}]_{0:T}) &=  \exp ^ {PL^2 ([\hat{F}, \hat{\theta}]_{0:T})}
\end{aligned} 
\label{eq:1_1}
\end{equation}

\noindent where $PL([\hat{F}, \hat{\theta}]_{0:T})$ is the physics law used to restrict the hierarchical structure of the generated data, which provides the additional information to the regularize the learning process from the small sample data. In this case, we use the Lagrange equation of motion \cite{zhao2022musculoskeletal} as the physics law, which is defined as follows.
\begin{equation}
\begin{aligned}
    PL([\hat{F}, \hat{\theta}]_{0:T}) = & \frac{1}{T} \sum_{t=1}^{T} (m(\hat{\theta}_t)\ddot{\hat{\theta}}_t + c(\hat{\theta}_t, \dot{\hat{\theta}}_t \\ & + g(\hat{\theta}_t) - \sum_{n=1}^{N} {\hat{F}^n_t})^2
\end{aligned} 
\label{eq:1_2}
\end{equation}

\noindent where $T$ is the number of time-steps, $N$ is the channels of the $\hat{F}$, $m(\hat{\theta}_t)$, $c(\hat{\theta}_t, \dot{\hat{\theta}}_t$, and $g(\hat{\theta}_t)$ denote mass matrix, the Centrifugal and Coriolis force, and the gravity, respectively \cite{zhang2022physics}. In this manner, the $G_{\sigma}$ will generate the structured outputs of $(\hat{F}, \hat{\theta})$. \par

The $Q_{D(\phi)}^{G(\sigma)}$ is computed by the $D(\phi)$ and interprets the physics constraint action values as the estimated probability of being physics real by $D(\phi)$. These physics constraint action values lead to the improvement of GAN model in physical extrapolation from the small training data. The $Q_{D(\phi)}^{G(\sigma)}$ can be formulated as:

\begin{equation}
\begin{aligned}
    Q_{D{\phi}}^{G{\sigma}}((G_{\sigma}(&sEMG_{0:T}), [F, \theta]_{0:T}) = \\
    & \mathbb{E}_{[\hat{F}, \hat{\theta}]_{0:T} \sim [F, \theta]_{0:T}} [\log D{\phi}([\hat{F}, \hat{\theta}]_{0:T})] + \\
    & \mathbb{E}_{[\hat{F}, \hat{\theta}]_{0:T} \sim G_{\sigma}(sEMG_{0:T}))}[\log (1-D{\phi}([\hat{F}, \hat{\theta}]_{0:T}))]
\end{aligned} 
\label{eq:2}
\end{equation}

For each epoch, once the new $R_G$ and $Q_{D(\phi)}^{G(\sigma)}$ has been obtained, the policy model $G(\sigma)$ will be updated following the gradient of the reward function as follows.

\begin{equation}
\begin{aligned}
    \nabla_{\sigma} J(\sigma) = & \mathbb{E}_{[\hat{F}, \hat{\theta}]_{0:T} \sim G_{\sigma}(sEMG_{0:T})} \sum \nabla_{\sigma} R_{G_{\sigma}}([\hat{F}, \hat{\theta}]_{0:T}|[F, \theta]_{0:T}) \\
    &\cdot Q^{G_{\sigma}}_{D_{\phi}} ([\hat{F}, \hat{\theta}]_{0:T}, [F, \theta]_{0:T})
\end{aligned} 
\label{eq:6}
\end{equation}

Using likelihood ratios, the unbiased estimation for Eq. \ref{eq:6} on one epoch can be described as follows.
    
\begin{equation}
\begin{aligned}
    &\nabla_{\sigma}J(\sigma) \simeq \frac{1}{T} \sum_{t=1}^{T} \sum_{y_t \in [\hat{F}, \hat{\theta}]_t} \nabla_{\sigma} R_{G_{\sigma}}(y_t|[F, \theta]_t) \cdot Q^{G_{\sigma}}_{D_{\phi}} (y_t, [F, \theta]_t)\\
    &=\frac{1}{T} \sum_{t=1}^T \sum_{y_t \in [\hat{F},\hat{\theta}]_t} G_\sigma(y_t|[F, \theta]_t) \nabla_\sigma \log G_\sigma(y_t|[F, \theta]_t) \\
    &\cdot Q^{G_{\sigma}}_{D_{\phi}}(y_t, [F, \theta]_t)
\end{aligned} 
\label{eq:7}
\end{equation}

The parameters of the policy model $G_{\sigma}$ can be updated as follows.

\begin{equation}
\begin{aligned}
    \sigma \leftarrow \sigma + \alpha \nabla_\sigma J(\sigma)
\end{aligned} 
\label{eq:8}
\end{equation}

\noindent where $\alpha \in \mathbb{R}$ is the learning rate. \par

\begin{algorithm}
\caption{Generative adversarial learning via physics-informed policy gradient}
\begin{algorithmic}[1]
\REQUIRE generator network $G_\sigma$; discriminator $D_\phi$; input multi-channel sEMG dataset $sEMG = \{X_{1:T}\}$; Inverse dynamics positive samples $Pos$
\STATE Initialize $G_\sigma$, $D_\phi$ with random weights $\sigma$, $\phi$.
\STATE Pre-train $G_\sigma$ using MLE on $sEMG$
\STATE $Pos \leftarrow G_\sigma$
\STATE Generate negative samples using $G_\sigma$ for training $D_\phi$
\STATE Pre-train $D_\phi$ via minimizing the cross entropy
\REPEAT
    \FOR{$G_\sigma$ training-steps}
        \STATE Generate $[\hat{F}, \hat{\theta}]$ time series via $G_\sigma$
        \FOR{t in 0:T}
            \STATE Compute $Q_{D{\phi}}^{G{\sigma}}$ by Eq. \ref{eq:2}
        \ENDFOR
        \STATE Update generator parameters via physics-informed reward Eq. \ref{eq:8}
    \ENDFOR
    \FOR{d-steps}
        \STATE Use current $G_\sigma$ to generate negative examples and combine them with given positive examples $Pos$
        \STATE Train discriminator $D_\phi$ for k epochs.
    \ENDFOR
    \STATE $\beta \leftarrow \sigma$
\UNTIL{GAN converges}
\end{algorithmic}
\end{algorithm}

To summarize, Algorithm 1 provides an in-depth look at our proposed GAN optimization via a physics-informed policy gradient. Initially, $G_\sigma$ is pre-trained on the training set $sEMG = \{X_{1:T}\}$ using the maximum likelihood estimation (MLE). And then, the $G_\sigma$ and $D_\phi$ undergo adversarial learning. As the $G_\sigma$ improves, the $D_\phi$ is routinely retrained to stay synchronized with the $G_\sigma$ improvement. We ensure balance by generating an equal number of negative samples for each training step as the positive samples. \par

\subsection{The generative network}
\label{meth:4}

\begin{figure}[]   
    \centering  
    \includegraphics[width=3.5in]{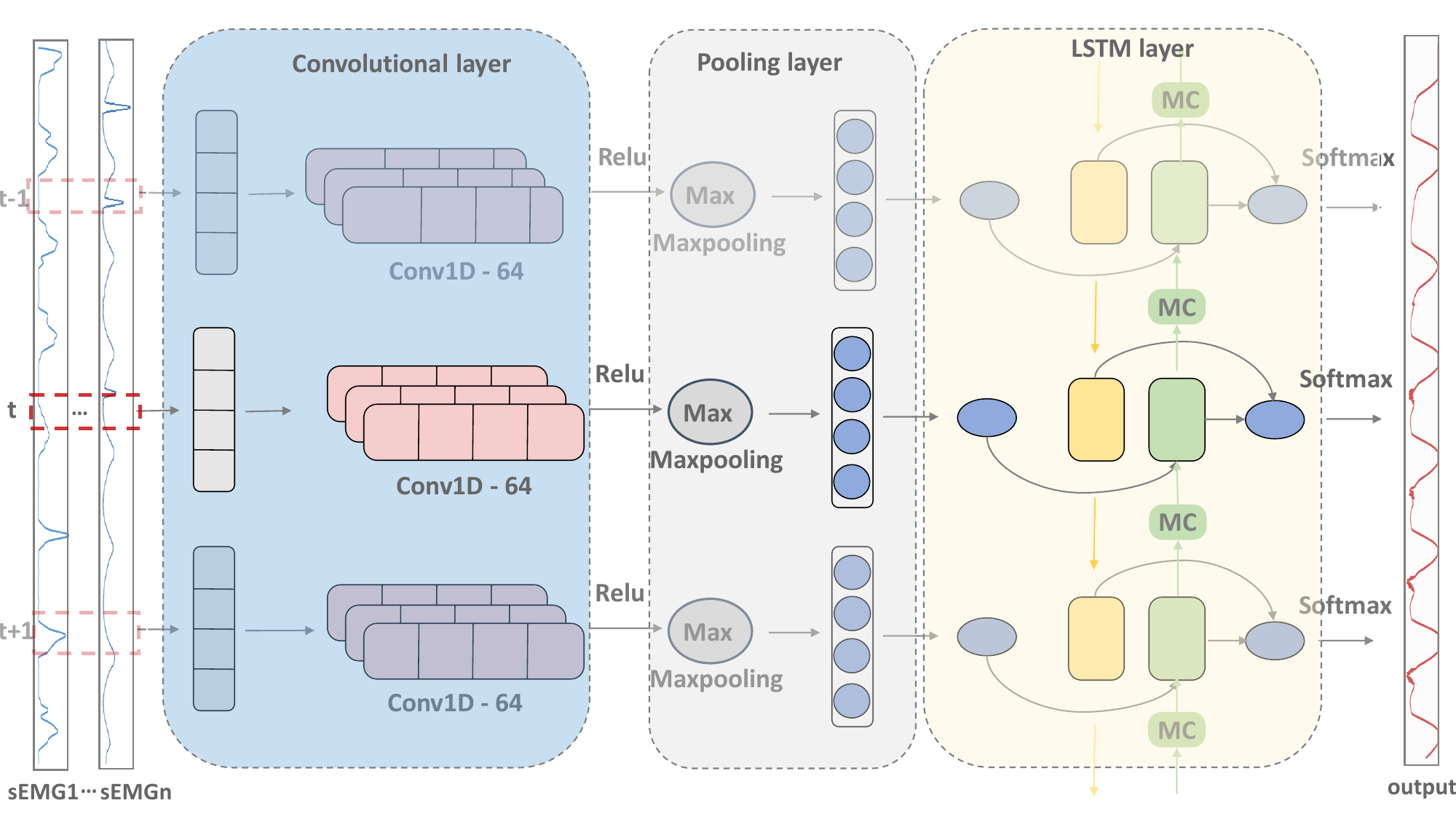}  
    \caption{The network architecture of the generator network in the proposed physics-informed reinforcement generative adversarial learning.}  
    \label{fig:1_G}  
\end{figure}

The proposed physics-informed low-shot learning method does not depend on the specific generative network architecture. In this study, considering the long-term temporal dependencies of the $F$ and $\theta$ sequences to the input multi-channel sEMG sequence, we employ the Long Short-Term Memory (LSTM) cells to our generative model \cite{han2019production}. The architecture of the generator network {$G$} is shown in Fig.\ref{fig:1_G}. It serves three functions: multi-channel sEMG feature extraction, residual learning with LSTM, and musculoskeletal tokens sequence generation. \par  

Firstly, for the multi-channel sEMG feature extraction, a 1-dimensional (1D) convolution filter with a $2 /times 1$ kernel is introduced to capture the multiple sEMG features at time step $t$. The extracted convolution features represent the hierarchical structures of the multi-channel sEMG. In this study, the convolution kernel is set to $1 \times b$ for a $b$-channel sEMG input. Considering the batch normalization (BN) layer would normalize the features and get rid of the range flexibility for upscaling features \cite{nah2017deep}, no BN layer is used here to avoid blurring the sEMG responses hidden in the extracted features. The max-pooling layer is used to combine the extracted sEMG features into a single neuron by using the maximum value from each convolution window. The max-pooling operation reduces the number of parameters and network computation costs and has the effect of adjusting over-fitting. \par

Secondly, the LSTM blocks are employed for residual learning of the time-series characteristics of the target musculoskeletal tokens. The LSTM layer is well suited for time-series sequence generation by addressing the explosive and vanishing gradient issues \cite{yu2017seqgan}. An LSTM block consists of a memory cell, an input gate, an output gate, and a forget gate, the detailed definitions of the components are described in \cite{nah2017deep}'s study. Specifically, in this study, in time step $t$, the memory cell remembers structured feature values over the previous $t-1$ intervals and the three gates regulate the flow of information into and out of the memory cell, which has a great preference for preserving long-term temporal structure characteristics by consolidating previous temporal correlations as memory units. Meanwhile, the high-level sEMG features extracted from the convolution layer represent the current multi-channel sEMG responses to muscle force and joint kinematics. The skip-connect of the memory cell and the high-level sEMG features not only represent extracted local kinetic invariances but also represent the temporal dynamics of the motions. \par 

It is noteworthy that the traditional LSTM layer only produces fitness between the current time step and the previous time steps. However, we expect the model also can pay insight into the resulting future outputs. In order to compute the action value for future physical fitness, a Monte Carlo ($MC$) search with a roll-out strategy is used to sample the unknown last $T-t$ time steps. and the $N$-time Monte Carlo search can be formulated as:

\begin{equation}
\begin{aligned}
    \{(F_{0:T}, \theta_{0:T})^1, ..., (F_{0:T}, \theta_{0:T})^N = MC(F_{0:t}, \theta_{0:t})\}
\end{aligned} 
\label{eq:3}
\end{equation}

Finally, the fully connected layers are used to generate the musculoskeletal tokens sequence over a motion period. The output of the LSTM unit is flattened to a feature vector and scaled to the muscle force $F$ and joint kinematics $\theta$. 

\subsection{The discriminative model}
\label{meth:5}

In this study, a $\phi$ parameterized discriminator network $D_{\phi}$ is built to guide the iterations of $G_{\sigma}$ from the small sample data. $D_{\phi}$ outputs a probability indicating the heterogeneity between $[\hat{F}, \hat{\theta}]$ and $[F, \theta]$. For this purpose, we employ a convolution neural network (CNN) \cite{wu2020review} as the discriminative model because of its successful applications in sequence classification. In this study, we concentrate on the situation where the discriminator estimates the likelihood of a completed $[\hat{F}, \hat{\theta}]$ time-series from the physical-law model (i.e. $ID$).\par

\begin{figure}[]   
    \centering  
    \includegraphics[width=3.5in]{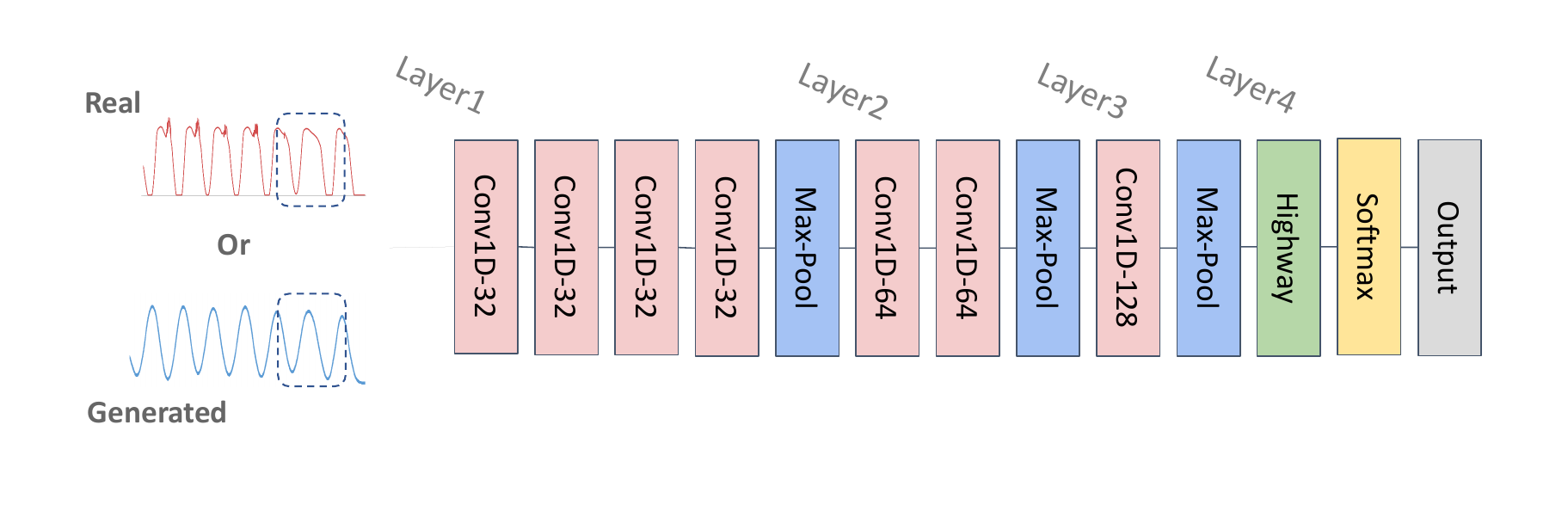}  
    \caption{The network architecture of the discriminative model in the proposed physics-informed reinforcement generative adversarial learning.}  
    \label{fig:1_D}  
\end{figure}

We first represent an input muscle force and joint kinematics time series $x_1,...,x_T$ as
\begin{equation}
\begin{aligned}
    E_{0:T} = [\hat{F}, \hat{\theta}]_0 \oplus [\hat{F}, \hat{\theta}]_2 \oplus ... \oplus [\hat{F}, \hat{\theta}]_T
\end{aligned} 
\label{eq:9}
\end{equation}
\noindent where, $x_t \in \mathbb{R}^b$ is the muscle force and joint kinematics in time-step $t$ and $\oplus$ is the concatenation operator to build the matrix $E_{1:T} \in \mathbb{R}^{T}$. Then the convolution operator is used to produce a new feature map:

\begin{equation}
\begin{aligned}
     c_i = \rho(w \odot E_{i:i+l-1} + b)
\end{aligned} 
\label{eq:10}
\end{equation}

\noindent where $\odot$ is the element-wise production, $b$ is a bias term and $\rho$ is a non-linear function. In this study, the discriminator, as shown in Fig.\ref{fig:1_D}, employs various numbers of kernels with different window sizes to extract different features from the input musculoskeletal sequence. And the max-pooling operation over the feature maps to reduce the number of parameters and network computation costs. In order to enhance the discrimination performance, a highway operator \cite{srivastava2015highway} based on the pooled feature maps is also employed in our discriminative model. Finally, a fully connected layer with softmax activation is used to output the estimation of the likelihood that the input sequence conforms to physical laws. 

\section{MATERIAL AND EXPERIMENTAL METHODS}
\label{Mat_and_ex}
In this study, we test our proposed method on two joint motion scenarios. The first one is the knee joint modeling from an open-access dataset of walking trials, and the second one is the wrist joint modeling from the self-collected dataset of wrist motions. \par

\subsection{Open-access dataset of walking trials}
The open-access dataset of walking trails is obtained from a real-world experiment reported in \cite{liu2008muscle}. This dataset involves six healthy participants with an average age of 12.9 $\pm$ 3.2 years and an average weight of 51.8 $\pm$ 19.1 Kg. Participants are instructed to walk at four distinct speeds, which include very slow (0.53 $\pm$ 0.1 m/s), slow (0.75 $\pm$ 0.1 m/s), free (1.15 $\pm$ 0.08 m/s), and fast (1.56 $\pm$ 0.21 m/s) speeds. The sEMG signals are captured from the $biceps femoris short head$ (BFS) and the $rectus femoris$ (RF) as they are the primary flexor and extensor of the knee joint. In this study, we normalize each gait cycle into 100 frames for model training and testing, and the original data for model extrapolation evaluation. In the model training and testing session, each walking trial sample is formatted into a source matrix that includes the time step, gait motion data, and enveloped sEMG signals. All of the samples from different participants are combined to create a comprehensive dataset for model training and testing. \par

\subsection{Self-collected dataset of wrist motions}
Our wrist motions experiment, approved by the MaPS and Engineering Joint Faculty Research Ethics Committee of the University of Leeds (MEEC 18-002), involved six participants with signed consent. Participants were instructed to keep their torso straight with their shoulder abducted at $90$ degrees and their elbow joint flexed at $90$ degrees. The VICON motion capture system is used to record continuous wrist flexion/extension motion. Joint motions are calculated using an upper limb model with 16 reflective markers with 250 Hz sampling rate. Concurrently, sEMG signals are captured from the primary wrist muscles (n = 1, 2,..., 5), including the $flexor carpi radialis$ (FCR), the $flexor carpi ulnaris$ (FCU), the $extensor carpi radialis longus$ (ECRL), the $extensor carpi radialis brevis$ (ECRB), and the $extensor carpi ulnaris$ (ECU) using Avanti Sensors (sampling rate is 2000 Hz). Electrodes are placed by palpation and their placement is validated by observing the signal during contraction before the experiment. The sEMG signals and motion data were synchronized and resampled at 1000 Hz. Each participant performed five repetitive trials with a three-minute break between trials to prevent muscle fatigue.\par

The recorded sEMG signals are pre-processed by a 20 Hz and 450 Hz band-pass filter, full rectification, and a 6 Hz low-pass filter. These signals are then normalized based on the maximum voluntary contraction recorded prior to the experiment, yielding the enveloped sEMG signals. We normalize each motion cycle into 156 frames for model training and testing, and the original data for model extrapolation evaluation. A total of 360 motion data are then combined to create a comprehensive dataset for model training and testing, and 6 motion data are used for model evaluation.

\subsection{Benchmark models and parameter settings}
To evaluate the performance and effectiveness of the proposed physics-informed policy gradient for low-shot generative adversarial learning, the benchmark models employ three representative methods, including physics-Informed convolutional neural network (PI-CNN) \cite{zhang2022physics} which represents the state-of-the-art deep learning based musculoskeletal modeling method, ML-ELM \cite{zhang2016multi} which represents the general musculoskeletal modeling method, and the vanilla GAN which represents the traditional GAN family without physical-law \cite{goodfellow2014distinguishability}.  \par


\subsection{Evaluation metrics}
The evaluation metrics include 1) the metrics for evaluating the quality of the generated samples including the information entropy associated peak signal-to-noise ratio (PSNR) \cite{solnik2008teager}, coefficient of Determination ($R^2$) \cite{kahl2016comparison}, root mean square error (RMSE) \cite{zhao2022computational}, Spearman's Rank Correlation Coefficient (SRCC) \cite{pan2022effect}, and 2) the metrics for evaluating the mode collapse of GANs, including 1) inception score (IS) \cite{shi2022latent}, and 2) Frechet inception distance (FID) \cite{jung2021internalized}.

\section{Results and discussion}
\label{result}
In this section, we evaluate the performance of the proposed physics-informed low-shot learning in the knee joint and wrist joint scenarios. We first carry out overall comparisons of the results from the proposed and benchmark methods. We also evaluate the model performance on small training data and handling mode collapse. Lastly, we investigate the robustness and generalization performance of the proposed method in intersession scenarios. The training of the proposed framework and benchmark methods was conducted using PyTorch on a workstation equipped with NVIDIA Quadro K4200 graphics cards and 256G RAM.\par


\subsection{Overall evaluation of the muscle force dynamics modeling}
\label{sec:3_OA}

\begin{figure}[]   
    \centering  
    \includegraphics[width=3.5in]{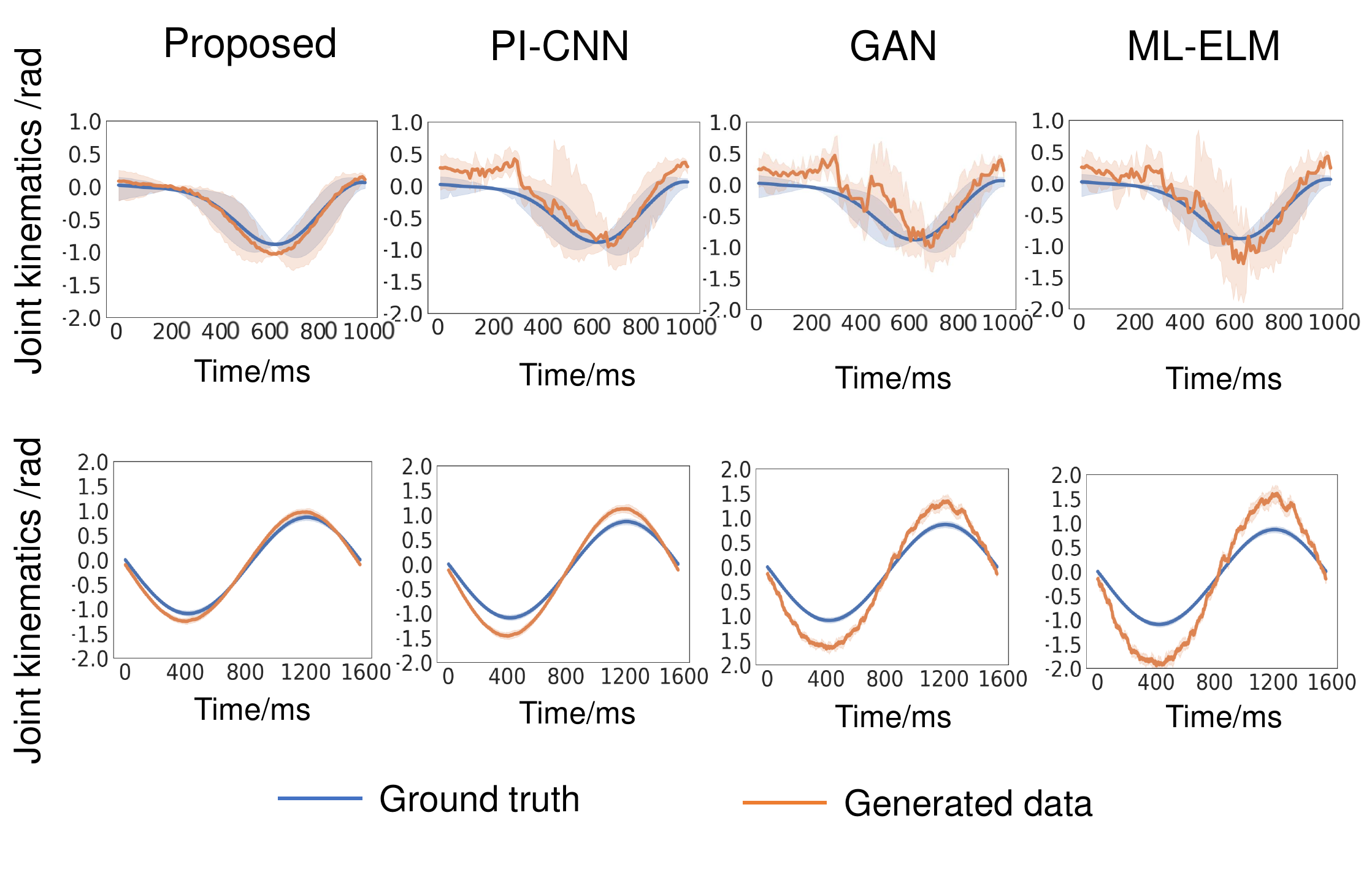}   
    \caption{Comparison of the average knee joint kinematics (the first row) and wrist joint kinematics (the second row) within one gait cycle between the ground truth and the generated data from the proposed and benchmark models. The shaded areas represent the mean $\pm$ one standard deviation of the kinematics. }  
    \label{fig:2_angle}  
\end{figure}

In this section, we first carry out overall comparisons between the proposed and benchmark methods on the test dataset. Fig. \ref{fig:2_angle} demonstrates the overall results of the joint kinematics generation in one motion circle from the proposed and benchmark methods for both the knee joint (the first row of Fig. \ref{fig:2_angle}) and wrist joint cases (the second row of Fig. \ref{fig:2_angle}). The average joint kinematics and standard deviation distribution from the proposed method align well with the ground truth in both the knee joint and wrist joint cases. These findings indicate the proposed model achieves the best performance among the benchmark models on the unbiased estimation of the joint kinematics.

\begin{figure}[]   
    \centering  
    \includegraphics[width=3.5in]{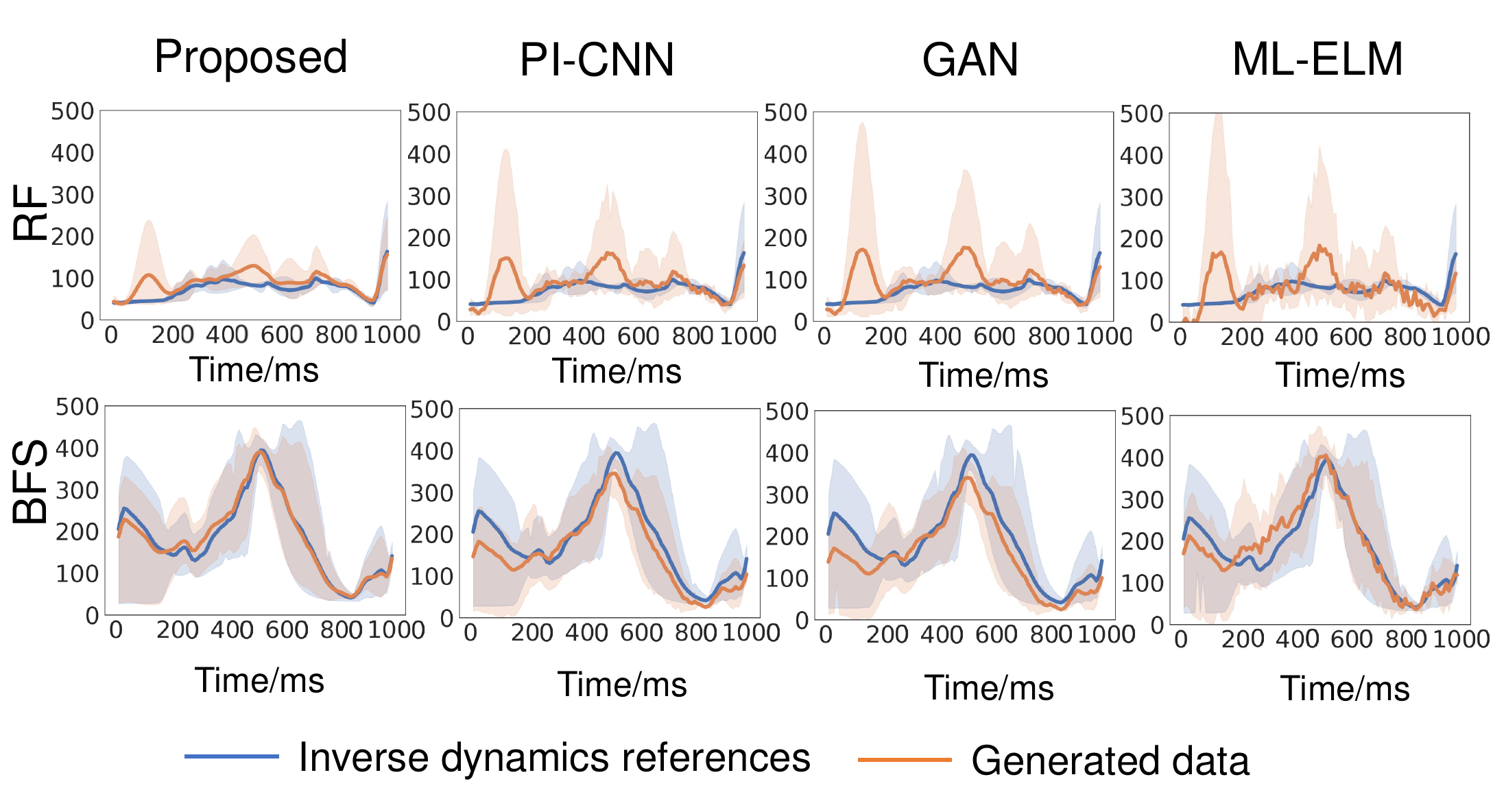}   
    \caption{Comparison of the average knee muscle force dynamics within one gait cycle between the real-target and the generated muscle force data from the proposed and benchmark models. The shaded areas represent the mean $\pm$ one standard deviation of the muscle force for BFS and RF. }  
    \label{fig:2_knee}  
\end{figure}

\begin{figure}[]   
    \centering  
    \includegraphics[width=3.5in]{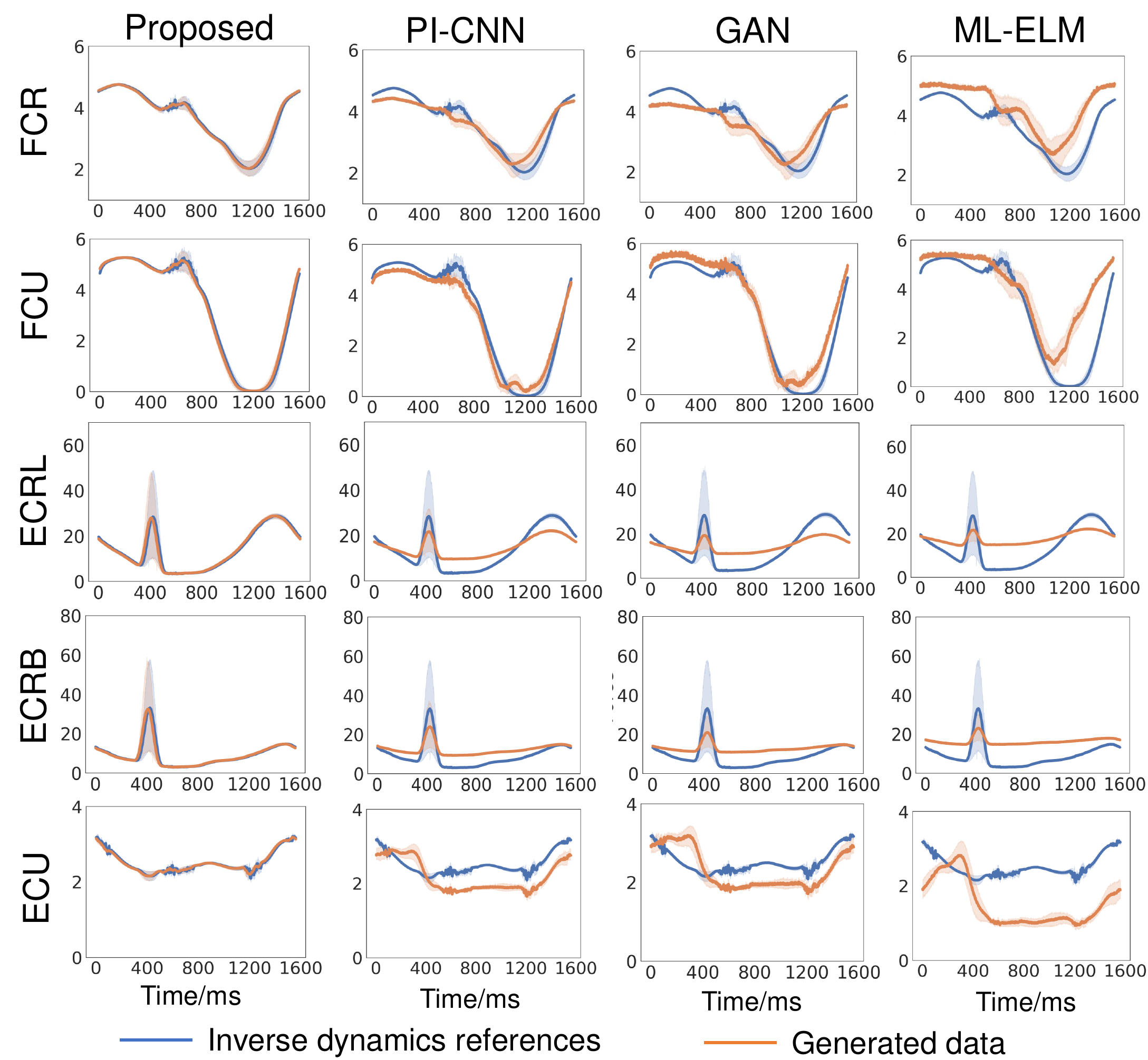}   
    \caption{Comparison of the average wrist muscle force dynamics within one motion cycle between the real-target and the generated muscle force data from 5-channel sEMG signal. The shaded areas represent the mean $\pm$ one standard deviation of the muscle force for FCR, FCU, ECRL, ECRB, and ECU. }  
    \label{fig:2}  
\end{figure}

Similarly, Fig. \ref{fig:2_knee} and Fig.\ref{fig:2} demonstrate the overall results of the muscle force estimations in one motion circle for both the knee joint (i.e. RF and BFS) and wrist joint (i.e. FCR, FCU, ECRL, ECRB, and ECU) cases, respectively. The average muscle forces estimated by the proposed method align well with the inverse dynamics, demonstrating the excellent multiple muscle tracking capability of the proposed model. In addition, the standard deviation distribution of the proposed model-generated muscle forces is perfectly consistent with the standard deviation distribution of the inverse dynamics-based references. These results indicate that the proposed model achieves the best performance among the benchmark models on the unbiased estimation of the muscle force from the multi-channel sEMG signals.    \par

To further assess the extrapolation performance quantitatively, we present detailed comparisons of the proposed and benchmark models on both of the test data and evaluation data. Table \ref{tab:1} and Table \ref{tab:2} respectively shows the results for the knee joint case and the wrist joint case. The results indicate that the proposed model performs best on both of the testing and evaluation data. Specifically, for model testing, the $PSNR$, $R^2$, $RMSE$, $SRCC$ of the proposed model are $15.57\%$, $6.22\%$, $28.08\%$, $7.2\%$ higher than that of the second best model (i.e. PI-CNN). For model evaluation, the $PSNR$, $R^2$, $RMSE$, $SRCC$ of the proposed model are $24.72\%$, $16.29\%$, $38.99\%$, $17.66\%$ higher than that of the second best model (i.e. GAN). In addition, because the evaluation data involve the original sEMG recordings, the comparison of the testing results and evaluation results indicates the model extrapolation from the experimental scenarios to real scenarios. The proposed model shows the best extrapolated estimation of muscle force and joint kinematics among the benchmark models, the results from the testing data and evaluation data is consistent. In contrast, the performance of the benchmark models show serious decline on evaluation data.


\begin{table}[]
\caption{The evaluation of the proposed and benchmark models on knee joint case with two-channels sEMG.}
\label{tab:1}
\centering
\resizebox{3.1in}{!}{
\begin{tabular}{cccccc}
\toprule
\multicolumn{1}{l}{}                   & \multicolumn{1}{l}{} & \multicolumn{4}{c}{Model test}       \\ \hline
\multicolumn{1}{l}{}                   & Methods              & PSNR     & R2      & RMSE    & SRCC  \\ \cline{2-6} 
\multirow{4}{*}{RF}                    & Proposed             & 91.91    & 0.88    & 11.32   & 0.92  \\
                                       & PI-CNN               & 77.45    & 0.84    & 19.64   & 0.85  \\
                                       & GAN                  & 75.54    & 0.82    & 18.25   & 0.81  \\
                                       & ML-ELM               & 59.94    & 0.76    & 25.62   & 0.72  \\ \cline{2-6} 
\multirow{4}{*}{BFS}                   & Proposed             & 93.45    & 0.93    & 11.93   & 0.93  \\
                                       & PI-CNN               & 76.93    & 0.87    & 19.21   & 0.83  \\
                                       & GAN                  & 76.17    & 0.85    & 18.35   & 0.79  \\
                                       & ML-ELM               & 62.66    & 0.78    & 26.43   & 0.73  \\ \cline{2-6} 
\multirow{4}{*}{$\theta$} & Proposed             & 34.79    & 0.91    & 5.73    & 0.92  \\
                                       & PI-CNN               & 30.16    & 0.84    & 5.97    & 0.89  \\
                                       & GAN                  & 30.89    & 0.88    & 6.57    & 0.85  \\
                                       & ML-ELM               & 21.33    & 0.75    & 11.25   & 0.73  \\ \hline
\multicolumn{1}{l}{}                   & \multicolumn{1}{l}{} & \multicolumn{4}{c}{Model evaluation} \\ \hline
\multirow{4}{*}{RF}                    & Proposed             & 88.89    & 0.82    & 11.21   & 0.83  \\
                                       & PI-CNN               & 58.91    & 0.59    & 24.17   & 0.6   \\
                                       & GAN                  & 68.72    & 0.7     & 26.51   & 0.69  \\
                                       & ML-ELM               & 46.79    & 0.53    & 28.75   & 0.5   \\ \cline{2-6} 
\multirow{4}{*}{BFS}                   & Proposed             & 91.84    & 0.91    & 11.91   & 0.84  \\
                                       & PI-CNN               & 58.19    & 0.61    & 23.58   & 0.58  \\
                                       & GAN                  & 69.26    & 0.72    & 25.79   & 0.67  \\
                                       & ML-ELM               & 49.21    & 0.55    & 38.98   & 0.51  \\ \cline{2-6} 
\multirow{4}{*}{$\theta$} & Proposed             & 34.89    & 0.92    & 5.45    & 0.91  \\
                                       & PI-CNN               & 23.43    & 0.59    & 8.3     & 0.62  \\
                                       & GAN                  & 28.27    & 0.75    & 7.89    & 0.72  \\
                                       & ML-ELM               & 17.19    & 0.53    & 18.44   & 0.51  \\\bottomrule
\end{tabular}
}
\end{table}

\begin{table*}[]
\caption{The evaluation of the proposed and benchmark models on wrist joint case with five-channels sEMG}
\label{tab:2}
\centering
\resizebox{5.5in}{!}{
\begin{tabular}{cccccccccccc}
\toprule
\multicolumn{12}{c}{Model test}                                                                                                  \\ \cline{2-12} 
                      & Methods  & PSNR  & R2   & RMSE  & SRCC &                        & Methods  & PSNR  & R2   & RMSE  & SRCC \\ \hline
\multirow{4}{*}{FCR}  & Proposed & 31.91 & 0.92 & 5.32  & 0.94 & \multirow{4}{*}{FCU}   & Proposed & 33.61 & 0.93 & 4.37  & 0.96 \\
                      & PI-CNN   & 27.45 & 0.84 & 9.64  & 0.83 &                        & PI-CNN   & 29.01 & 0.86 & 10.43 & 0.83 \\
                      & GAN      & 25.54 & 0.86 & 8.25  & 0.81 &                        & GAN      & 25.27 & 0.88 & 8.6   & 0.79 \\
                      & ML-ELM   & 19.94 & 0.74 & 15.62 & 0.72 &                        & ML-ELM   & 18.42 & 0.76 & 14.95 & 0.73 \\ \cline{2-6} \cline{8-12} 
\multirow{4}{*}{ECRL} & Proposed & 84.21 & 0.95 & 14.68 & 0.94 & \multirow{4}{*}{ECRB}  & Proposed & 82.93 & 0.95 & 14.78 & 0.97 \\
                      & PI-CNN   & 79.4  & 0.84 & 25.08 & 0.83 &                        & PI-CNN   & 79.75 & 0.88 & 24.32 & 0.81 \\
                      & GAN      & 61.54 & 0.9  & 24.55 & 0.82 &                        & GAN      & 59.71 & 0.91 & 24.62 & 0.79 \\
                      & ML-ELM   & 57.76 & 0.77 & 42.41 & 0.76 &                        & ML-ELM   & 57.4  & 0.78 & 41.82 & 0.77 \\ \cline{2-6} \cline{8-12} 
\multirow{4}{*}{ECU}  & Proposed & 30.81 & 0.92 & 5.14  & 0.92 & \multirow{4}{*}{theta} & Proposed & 34.32 & 0.97 & 3.75  & 0.96 \\
                      & PI-CNN   & 30.31 & 0.84 & 10.06 & 0.82 &                        & PI-CNN   & 29.94 & 0.84 & 4.63  & 0.88 \\
                      & GAN      & 28.06 & 0.87 & 7.92  & 0.8  &                        & GAN      & 30.34 & 0.86 & 4.51  & 0.85 \\
                      & ML-ELM   & 19.85 & 0.75 & 14.72 & 0.71 &                        & ML-ELM   & 21.15 & 0.76 & 9.62  & 0.74 \\ \hline
\multicolumn{12}{c}{Model evaluation}                                                                                            \\ \hline
\multirow{4}{*}{FCR}  & Proposed & 29.96 & 0.87 & 5.05  & 0.89 & \multirow{4}{*}{FCU}   & Proposed & 31.35 & 0.88 & 4.15  & 0.91 \\
                      & PI-CNN   & 20.49 & 0.63 & 10.23 & 0.62 &                        & PI-CNN   & 21.75 & 0.65 & 9.82  & 0.62 \\
                      & GAN      & 22.43 & 0.77 & 11.43 & 0.73 &                        & GAN      & 21.8  & 0.79 & 12.74 & 0.71 \\
                      & ML-ELM   & 14.09 & 0.56 & 19.72 & 0.54 &                        & ML-ELM   & 13.57 & 0.57 & 21.21 & 0.55 \\ \cline{2-6} \cline{8-12} 
\multirow{4}{*}{ECRL} & Proposed & 79.76 & 0.9  & 13.95 & 0.89 & \multirow{4}{*}{ECRB}  & Proposed & 78.33 & 0.9  & 14.04 & 0.92 \\
                      & PI-CNN   & 58.65 & 0.63 & 28.81 & 0.62 &                        & PI-CNN   & 59.81 & 0.66 & 28.24 & 0.61 \\
                      & GAN      & 54.52 & 0.81 & 32.1  & 0.74 &                        & GAN      & 53.7  & 0.82 & 24.11 & 0.71 \\
                      & ML-ELM   & 42.45 & 0.58 & 39.81 & 0.57 &                        & ML-ELM   & 42.3  & 0.59 & 51.37 & 0.58 \\ \cline{2-6} \cline{8-12} 
\multirow{4}{*}{ECU}  & Proposed & 28.64 & 0.87 & 4.88  & 0.87 & \multirow{4}{*}{theta} & Proposed & 31.75 & 0.92 & 3.56  & 0.91 \\
                      & PI-CNN   & 22.29 & 0.63 & 10.55 & 0.62 &                        & PI-CNN   & 21.73 & 0.63 & 6.47  & 0.66 \\
                      & GAN      & 24.41 & 0.78 & 11.13 & 0.72 &                        & GAN      & 25.58 & 0.77 & 8.06  & 0.77 \\
                      & ML-ELM   & 14.28 & 0.56 & 16.04 & 0.53 &                        & ML-ELM   & 15.43 & 0.57 & 11.22 & 0.56 \\\bottomrule
\end{tabular}
}
\end{table*}

\subsection{Evaluation of low-shot learning}
\label{sec:3_ls}
The proposed physics-informed policy gradient incorporates the temporal relationship of the muscle force and joint kinematics dynamics from the Lagrange motion equation, resulting in an improved kinetics estimation from the low-shot samples. Initially, the physical information is used to constrain the model reward accumulated following the periodic multi-channel sEMG signals. And then, the accumulative reward is used to guide the Monte Carlo search to generate the unbiased estimation of muscle force and joint kinematics dynamics. \par 
To quantitatively assess the effectiveness of the proposed method on low-shot learning, we firstly regard the modeling results shown in Table \ref{tab:1} and Table \ref{tab:2} as the baselines that represent the optimal performance of the proposed and benchmark models, and then we train the models with different training sample sizes for $1500$ epochs as low-shot learning learning. The percentages of the low-shot learning learning results and the baseline joint kinematics modeling results, denote as $P-PSNR$, $P-R^2$, $P-RMSE$, and $P-SRCC$, are used as the evaluation metrics to describe what percentage of the performance of the baseline models can be achieved with the new models. \par

The evaluation of the low-shot learning of the proposed and benchmark models on the knee joint and wrist joint kinematics modeling is shown in Table \ref{tab:3}. It is obvious that the proposed model with a physics-informed policy gradient outperforms all of the benchmark models in low-shot learning. The 10-shot learning is able to achieve over $80\%$ baseline performance in terms of $PSNR$, $R^2$, $RMSE$, and $SRCC$. In comparison, the PINN and GAN models achieving a similar modeling performance require at least 80-shot learning. Therefore, it can be inferred that the proposed physics-informed policy gradient relies heavily on the physical representations and temporal structural characteristics of the training data, rather than the quantity of the data. This is encouraging as it suggests that the proposed method facilitates the applications of deep learning in biomechanical engineering from the general issue of limited sample size.\par

\begin{table*}[]
\caption{Evaluation of the low-shot learning performance of the proposed and benchmark models on joint kinematics modeling. The $P-PSNR$, $P-R^2$, $P-RMSE$, and $P-SRCC$ respectively represent the $SNR$, $R^2$, $RMSE$, and $SRCC$ of the $n$-shot learning as a percentage of the validation metrics of the best joint kinematics results report in Table.\ref{tab:1} and Table \ref{tab:2}.}
\label{tab:3}
\centering
\resizebox{5.5in}{!}{
\begin{tabular}{cccccccccc}
\toprule
                          &          & \multicolumn{4}{c}{Knee joint case}         & \multicolumn{4}{c}{Wrist joint case}        \\ \cline{3-10} 
                          &          & P-PNSR & P-$R^2$ & P-RMSE & P-SRCC & P-PNSR & P-$R^2$ & P-RMSE & P-SRCC \\ \midrule
\multirow{7}{*}{Proposed} & 1-shot   & 75\%      & 74\%    & 76\%      & 75\%      & 76\%      & 73\%    & 77\%      & 75\%      \\
                          & 10-shot  & 83\%      & 82\%    & 84\%      & 87\%      & 82\%      & 81\%    & 84\%      & 88\%      \\
                          & 20-shot  & 86\%      & 84\%    & 86\%      & 86\%      & 87\%      & 86\%    & 88\%      & 84\%      \\
                          & 40-shot  & 92\%      & 91\%    & 92\%      & 91\%      & 93\%      & 91\%    & 93\%      & 94\%      \\
                          & 60-shot  & 94\%      & 94\%    & 92\%      & 94\%      & 96\%      & 97\%    & 93\%      & 93\%      \\
                          & 80-shot  & 93\%      & 94\%    & 95\%      & 94\%      & 92\%      & 93\%    & 97\%      & 94\%      \\
                          & 100-shot & 95\%      & 94\%    & 93\%      & 93\%      & 96\%      & 94\%    & 93\%      & 96\%      \\ \midrule
                          &          & P-PNSR & P-$R^2$ & P-RMSE & P-SRCC & P-PNSR & P-$R^2$ & P-RMSE & P-SRCC \\ \midrule
\multirow{7}{*}{PINN}     & 1-shot   & 41\%       & 41\%      & 41\%      & 39\%       & 42\%       & 42\%      & 44\%       & 39\%       \\
                          & 10-shot  & 44\%       & 42\%     & 44\%       & 44\%       & 46\%       & 42\%     & 45\%       & 47\%       \\
                          & 20-shot  & 68\%       & 69\%     & 72\%       & 73\%       & 69\%       & 70\%      & 72\%       & 76\%       \\
                          & 40-shot  & 76\%       & 76\%     & 77\%       & 79\%       & 77\%       & 78\%     & 8\%        & 78\%       \\
                          & 60-shot  & 79\%       & 77\%     & 76\%       & 75\%       & 78\%       & 77\%     & 76\%       & 76\%       \\
                          & 80-shot  & 82\%       & 83\%     & 84\%       & 85\%       & 81\%       & 86\%     & 83\%       & 84\%       \\
                          & 100-shot & 84\%       & 87\%     & 85\%       & 87\%       & 85\%       & 88\%     & 85\%       & 86\%       \\ \midrule
                          &          & P-PNSR & P-$R^2$ & P-RMSE & P-SRCC & P-PNSR & P-$R^2$ & P-RMSE & P-SRCC \\ \midrule
\multirow{7}{*}{GAN}      & 1-shot   & 46\%       & 44\%     & 47\%       & 49\%       & 45\%       & 45\%     & 48\%       & 51\%       \\
                          & 10-shot  & 45\%       & 45\%     & 45\%       & 47\%       & 48\%       & 46\%     & 44\%       & 48\%       \\
                          & 20-shot  & 66\%       & 69\%     & 7\%        & 73\%       & 67\%      & 71\%     & 70\%        & 73\%      \\
                          & 40-shot  & 72\%       & 73\%     & 74\%       & 74\%       & 74\%       & 72\%     & 72\%       & 76\%       \\
                          & 60-shot  & 79\%       & 78\%     & 81\%        & 81\%       & 78\%       & 78\%     & 80\%        & 8\%        \\
                          & 80-shot  & 81\%       & 83\%     & 85\%       & 85\%       & 79\%       & 83\%     & 86\%       & 84\%       \\
                          & 100-shot & 84\%       & 86\%     & 87\%       & 89\%       & 86\%       & 87\%     & 86\%       & 91\%       \\ \midrule
                          &          & P-PNSR & P-$R^2$ & P-RMSE & P-SRCC & P-PNSR & P-$R^2$ & P-RMSE & P-SRCC \\ \midrule
\multirow{7}{*}{ML-ELM}   & 1-shot   & 36\%       & 35\%     & 37\%       & 38\%       & 34\%       & 37\%     & 36\%       & 37\%       \\
                          & 10-shot  & 38\%       & 44\%      & 45\%        & 39\%       & 39\%       & 39\%     & 42\%       & 38\%       \\
                          & 20-shot  & 57\%       & 56\%     & 55\%       & 54\%       & 59\%       & 56\%    & 57\%       & 55\%       \\
                          & 40-shot  & 62\%       & 62\%     & 65\%        & 59\%       & 65\%        & 61\%    & 68\%        & 58\%      \\
                          & 60-shot  & 66\%       & 65\%     & 67\%       & 66\%       & 65\%       & 64\%     & 66\%       & 67\%       \\
                          & 80-shot  & 75\%       & 73\%     & 72\%       & 74\%       & 77\%       & 74\%     & 71\%       & 74\%       \\
                          & 100-shot & 78\%       & 79\%     & 78\%        & 81\%       & 78\%       & 82\%     & 81\%       & 82\%       \\ \bottomrule
\end{tabular}
}
\end{table*}
 
\subsection{Mode collapse evaluation}
\label{sec:3_mc}
Mathematically, the generative model is easy to find a biased estimation caused by mode collapse, which leads to the generated samples only being located in the partial real distribution where it can fool the discriminative model and ignore other modes of real distribution during the adversarial learning. To handle this issue, the proposed physics-informed policy gradient alleviates the random noises and makes the generated feature sequence governed by the physics law, which facilitates the estimation of compound kinematics patterns and achieves the unbiased estimation of kinematics generation.\par

In order to evaluate the performance of the proposed method on alleviating the mode collapse, we test and compare the proposed model with the benchmark model from two aspects: 1) a quantitative evaluation of the diversity of the generated motions, based on the distance-derived IS and FID metrics; and 2) a monotonicity assessment on the generator iterations during the network training process; and 3) visualization of the distributions of the real and the generated motion samples.
Firstly, the quantitative evaluation for the diversity of the generated motions is conducted on the testing dataset. The higher IS and lower FID indicate the better diversity of the generated super-resolution HSIs, which further indicates the alleviation of mode collapse. \par

The results demonstrated in Table \ref{table:mode_collapse} show the proposed model outperforms the competitors in terms of the IS and FID measurements for both the knee joint and wrist joint motion generation. In addition, the benchmark GAN model, with the network architecture as same as the proposed model, is $19.11\%$ higher in IS, and $14.23\%$ lower in FID than the proposed model. These findings suggest that the proposed physics-informed policy gradient optimization approach has great performance in alleviating the mode collapse during adversarial learning.  \par

\begin{table}[]
\caption{The Comparison of Inception Scores (IS) and Frechet Inception Distances (FID) of the joint kinematics generated from the proposed and benchmark models on the model test datasets.}
\label{table:mode_collapse} 
\centering
\resizebox{3in}{!}{
\begin{tabular}{ccccc}
\toprule
         & \multicolumn{2}{c}{Knee joint case} & \multicolumn{2}{c}{Wrist joint case} \\ \cline{2-5} 
         & IS               & FID              & IS                & FID              \\ \hline
Proposed & 10.59            & 79.83            & 6.79             & 57.05            \\
PINN     & 13.11            & 78.54            & 8.43             & 46.13            \\
GAN      & 12.68            & 71.13            & 8.03             & 48.8             \\
ML-ELM   & 15.39            & 74.42            & 12.5              & 41.95            \\ \bottomrule
\end{tabular}
}
\end{table}

\begin{figure}[]   
    \centering  
    \includegraphics[width=3.5in]{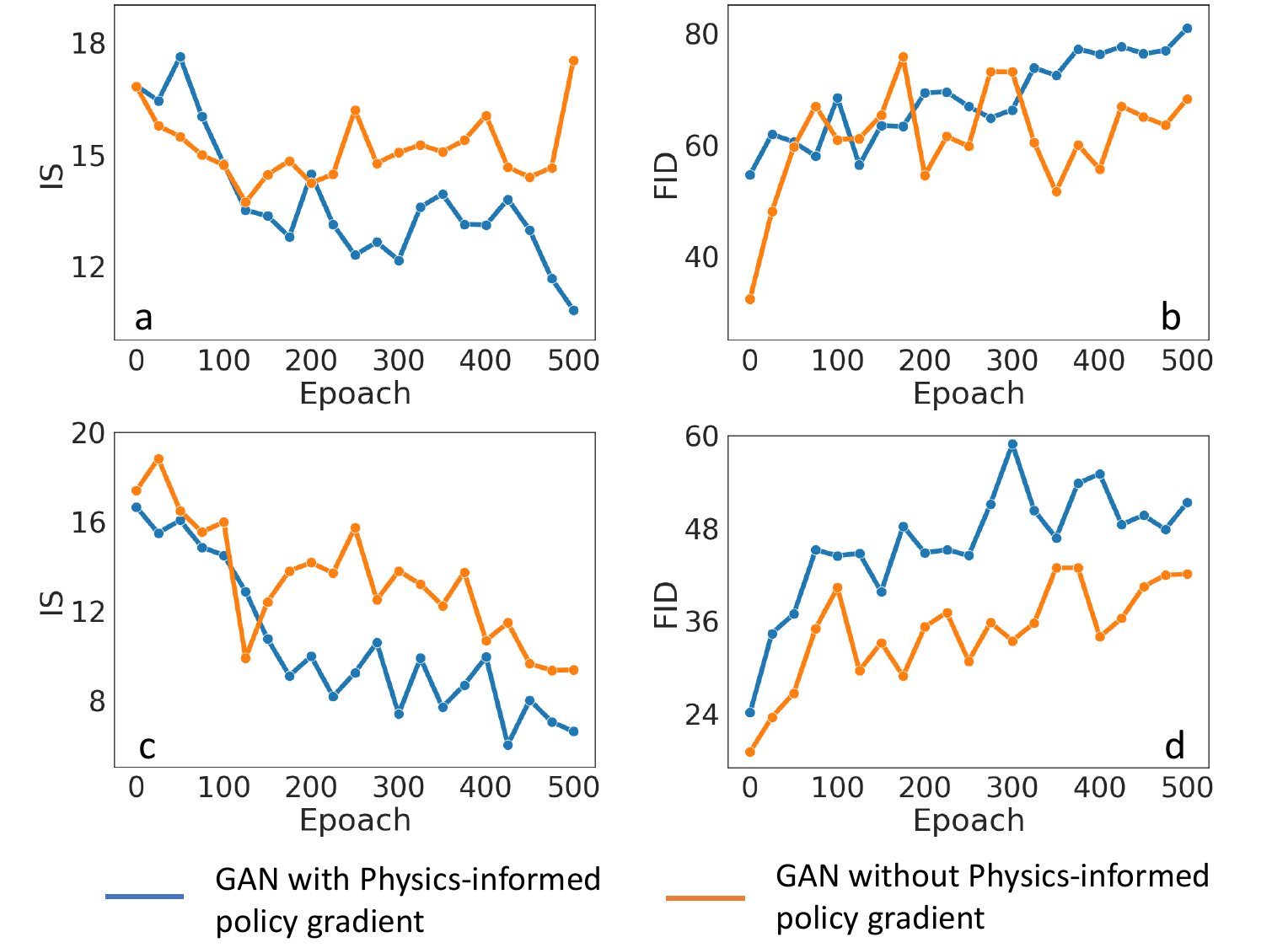}   
    \caption{Changes of IS and FID scores of the generated joint kinematics during the first 500 iterations of the GAN model using the proposed physics-informed policy gradient and the typical GAN without using the physics-informed policy gradient. The test is conducted on knee joint cases (a) and (b) and wrist joint cases (c) and (d), respectively.}  
    \label{fig:4_mode}  
\end{figure}

Secondly, in order to further explore the performance of the proposed physics-informed policy gradient on the mode collapse issue, we compare the generator iterations of the same GAN architectures with and without the physics-informed policy gradient (Fig. \ref{fig:4_mode}). The IS and FID curves from the GAN with the proposed physics-informed policy gradient are more monotonous than the GAN without the physics-informed policy gradient, along with the increase of iteration number. Thus, the curves of IS from the proposed physics-informed policy gradient steadily increase and the curves of FID steadily decrease for both knee joint (\ref{fig:4_mode}a and b) and wrist joint (\ref{fig:4_mode}c and d) cases.  \par

\subsection{Model application on intra-session scenario}
\label{sec:3_is1}
\begin{figure}[]   
    \centering  
    \includegraphics[width=3.5in]{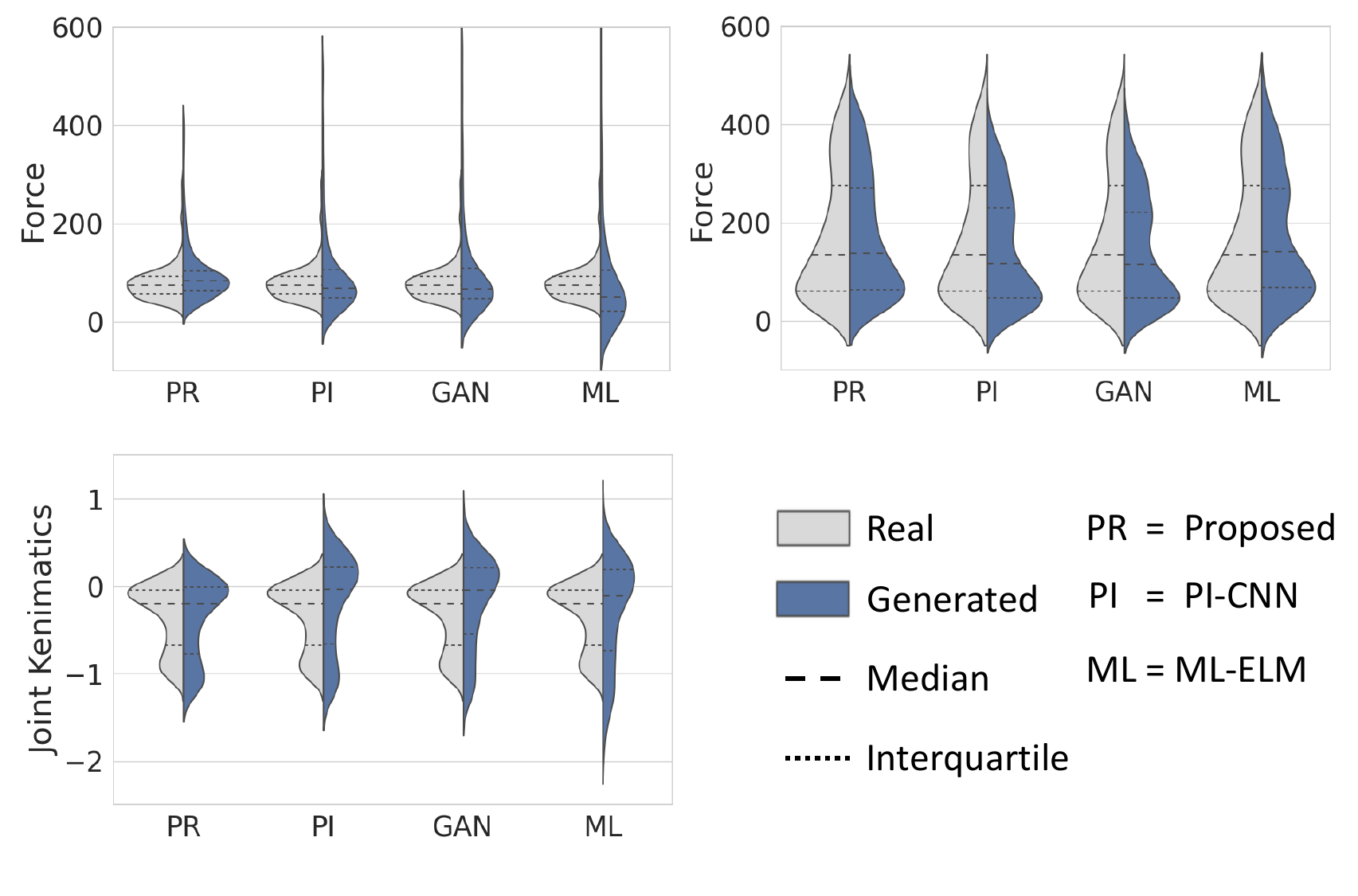}   
    \caption{Robustness evaluation of the proposed model (PR), PI-CNN (PI), GAN, and ML-ELM (ML) on the intra-session scenario.}  
    \label{fig:6_intra}  
\end{figure}

In musculoskeletal modeling, the intra-session scenario is regarded as the multiple sets of motions that occur within the same session. To test the robustness of the proposed model in the intra-session scenario, we use the knee joint data with different walking speeds for one subject as the intra-session evaluation dataset. The muscle force and joint kinematics modeling results, as shown in Fig. \ref{fig:6_intra}, indicate that the proposed framework performs best among the baseline methods. Importantly, the median and interquartile values of the proposed model with physics-informed policy gradient remain consistent with the real data across different walking speeds. In comparison, the median and quartiles of the baseline methods, such as the GAN model without using the physics-informed policy gradient, show significant inconsistencies with the real data, indicating a declined performance in the intra-session scenario due to the variability in walking speeds. These findings suggest that the model optimized by the proposed physics-informed policy gradient has great robustness in intra-session scenarios. 

\subsection{Model application on inter-session scenario}
\label{sec:3_is2}
\begin{figure}[]   
    \centering  
    \includegraphics[width=3.5in]{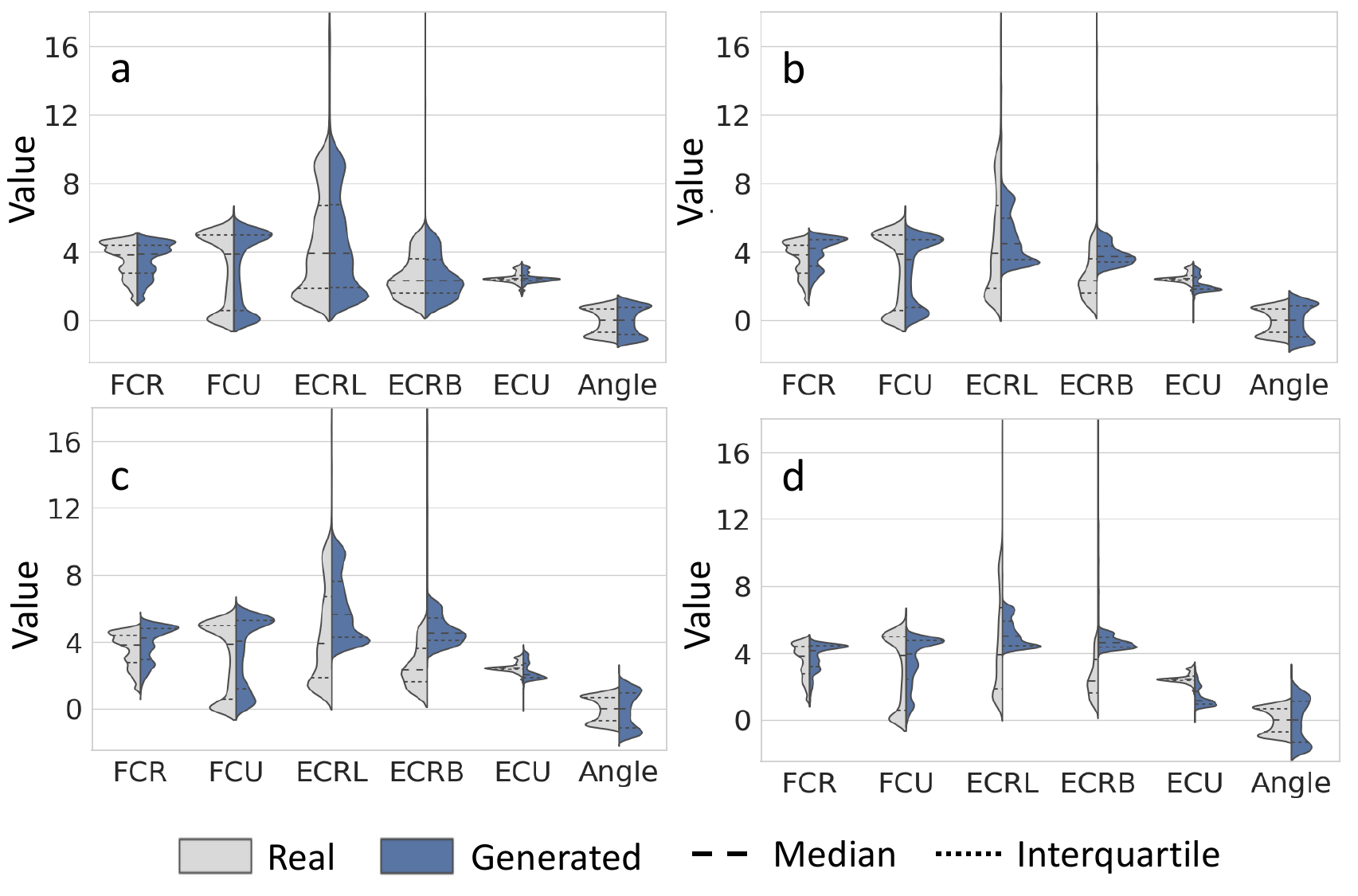}   
    \caption{Robustness evaluation of the a) proposed model, b) PI-CNN, c) GAN, and d) ML-ELM (ML) on the inter-session scenario.}  
    \label{fig:6_inter}  
\end{figure}

The inter-session scenario generally refers to a situation where motion data are collected across multiple sessions. To test the robustness of the proposed model in the inter-session scenario, we use the wrist joint data with different subjects as the evaluation dataset. The muscle force and joint kinematics modeling results, as shown in Fig. \ref{fig:6_inter}, indicate that the proposed framework performs best on the musculoskeletal modeling among the baseline methods. Specifically, the median and interquartile values of the proposed model with physics-informed policy gradient remain consistent with the real data across different subjects. In comparison, the baseline methods, such as the GAN model without using the physics-informed policy gradient, show a declined performance in the inter-session scenario due to the variability in walking speeds. These findings suggest that the model optimized by the proposed physics-informed policy gradient has great robustness in inter-session scenarios. 

\section{Conclusion}
\label{conclusion}
This paper develops a physics-informed low-shot learning method, which seamlessly integrates the Lagrange equation of motion and inverse dynamic muscle model into the adversarial learning process, to train the generative network for the unbiased estimation of the muscle force and joint kinematics from the small size sEMG time series. Specifically, the Lagrange equation of motion is introduced as physical constraint, which facilitates the generator to estimate the muscle force and joint kinematics with more temporal structural representations. Meanwhile, the physics-informed policy gradient rewards the physical consistency of the generated muscle force and joint kinematics and the inverse dynamics-based references, which improve the extrapolation performance of the generative network. Comprehensive experiments on the knee joints and wrist joints indicate the feasibility of the proposed method. The resultant findings suggest that the proposed method performs well in handling the mode collapse issue on the small sample data, and the estimations of the muscle forces and joint kinematics are unbiased compared to the physics-based inverse dynamics. These findings suggest that the proposed method may reduce the gaps between laboratory prototypes and clinical applications. However, it is worth noting that the physics reference (i.e. the inverse dynamics for this study) plays an important role in constraining the physics representation of the generated samples. Therefore, the choice of physics module may vary when the proposed approach is extended to other application cases.\par

Going forward, we plan to delve deeper into the properties of the physics-informed deep learning framework in the context of sEMG-based musculoskeletal modeling. We aim to investigate the potential of the low-shot learning-based model on the continuous and simultaneous estimation of multiple joint kinematic chains from sEMG signals. We also plan to adjust the compositions of the proposed method to cater to different application scenarios. Furthermore, we intend to evaluate the reliability and accuracy of the proposed framework through more complex movements.

\bibliographystyle{unsrtnat}
\bibliography{ref}  






\end{document}